\documentclass[a4paper]{jpconf}
\usepackage{graphicx}
\RequirePackage{lscape}
\begin{document}
\title{Review on dark matter searches}

\author{Susana Cebri\'an}

\address{Centro de Astropart\'iculas y F\'isica de Altas Energ\'ias (CAPA), Universidad de Zaragoza, 50009 Zaragoza, Spain
\\ Laboratorio Subterr\'aneo de Canfranc, 22880 Canfranc Estaci\'on, Huesca, Spain}

\ead{scebrian@unizar.es}

\begin{abstract} 
Dark matter particles populating our galactic halo could be directly detected by measuring their scattering off target nuclei or electrons in a suitable detector. As this interaction is expected to occur with very low probability and would generate very small energy deposits, the detection is challenging; the possible identification of distinctive signatures (like an annual modulation in the interaction rates or directionality) to assign a dark matter origin to a possible observation is being considered. Here, the physics case of different dark matter direct detection experiments will be presented and the different and complementary techniques which are being applied or considered will be discussed, summarizing their features and latest results obtained. Special focus will be made on TPC-related projects; experiments using noble liquids have presently a leading role to constrain interaction cross sections of a wide range of dark matter candidates and gaseous detectors are very promising to explore specifically low mass dark matter as well as to measure directionality.

\end{abstract}

\section{Introduction}

There is an overwhelming evidence of the existence of dark matter (DM) from observations at different scales, from galaxies to the whole Universe, supporting that a large fraction of the mass and energy budget is not explained within the standard cosmological model~\cite{historydm}. Proposed DM candidates include non-zero-mass particles having a very low interaction probability with baryonic matter, like the so-called Weakly Interacting Massive Particles (WIMPs). Different complementary strategies for detection of DM particles which can be pervading the galactic halo are being considered, like production at colliders and the indirect detection through standard particles produced by their annihiliation. Here, the focus is on the direct detection registering typically the elastic scattering of WIMPs off target nuclei~\cite{schumann2019,appecdm}; it is challenging as it produces a rare signal, concentrated at very low energies with a continuum energy spectrum which appears entangled with background. Therefore, ultra low background conditions and very low energy threshold are a must and the identification of distinctive signatures would be extremely helpful to assign a DM origin to a possible observation. Different physics cases investigated in this context will be discussed in Sec.~\ref{secpc} while the different techniques applied and latest results released will be summarized in Sec.~\ref{sectech}.

\section{Physics cases}
\label{secpc}

Many of the experiments attempting the direct detection of DM are searching for an excess of events over background, considering different possibilities: for the interactions, being Spin-Independent (SI) or Spin-Dependent (SD); the scattering, either on nuclei or electrons producing Nuclear or Electronic Recoils (NR/ER); and the masses of the DM particles, existing well-motivated candidates in the range from eV to TeV. Indeed, a lot of effort is being devoted in the last years to probe low mass DM below the GeV scale~\cite{whitepaper}; this additionally requires lighter targets (to keep kinematic matching between WIMPs and nuclei) and lower threshold (to detect smaller energy deposits) and has forced to consider different search channels. As light WIMPs cannot transfer sufficient momentum to generate detectable NR, absorption or scattering off by electrons (ER) are being taken into account too. But there are also experiments specifically  focused on the observation of distinctive DM signatures, like an annual modulation of the interaction rates and the directionality of the signal. In this section, these two signatures and the proposed Migdal effect allowing to enhance sensitivity to low mass DM will be briefly presented.

\subsection{Annual modulation}

The movement of the Earth around the Sun makes the relative velocity between DM particles in the galactic halo and a detector follow a cosine variation producing a time modulation in the DM interaction rate $S$, which for each energy bin $k$ is typically expressed as the sum of a constant term, $S_{0,k}$, and a modulated one with amplitude $S_{m,k}$,
\begin{equation}
S_{k}=S_{0,k}+S_{m,k}\cos(\omega(t-t_{0})).
\end{equation}
For a locally isotropic DM halo, the modulation has a one year period (related to $\omega$) and a well defined phase $t_{0}$, being the maximum rate around June, 2$^{nd}$. The modulation should be weak (at the level of a few percent variation), only noticeable in the low energy region and with a phase reversal which can produce a negative modulation amplitude at very low energies~\cite{drukier,freese1988,freese2013}. No background component is known to mimic the effect and therefore its observation would be a distinctive signature of DM interaction.

\subsection{Directionality}

The average direction of the WIMP wind through the solar system comes from the constellation of Cygnus, as the Sun is moving around the Galactic center. A measurement of the track direction of NR could be used then to distinguish a DM signal from background events (expected to be uniformly distributed) and to prove the galactic origin of a possible signal~\cite{spergel,ahlen,mayet}. The reconstruction of tracks is not easy since they are very short for keV scale NR: $\sim$1~mm in gas, $\sim$0.1~$\mu$m in solids. It would be desirable to register direction (axis, sense) or at least a head-tail asymmetry by measuring the relative energy loss along the track. A daily modulation of the WIMP direction due to the rotation of Earth has also been proposed. 

\subsection{Migdal effect}

The Migdal effect is an Atomic Physics phenomenon that leads to the emission of a bound-state electron from atomic or molecular systems when the atomic nucleus is suddenly perturbed, which has been observed for radioactive $\alpha$ and $\beta$ decays. Therefore, the WIMP-nucleus interaction could lead to atomic excitation or ionization of the recoiling atom~\cite{pradler,ibe,dolan}. Although there is no experimental evidence of the Migdal effect for NR yet, it has been already considered by several collaborations to explore sub-GeV mass DM candidates as, for low mass DM, this additional signal is above threshold (unlike the NR alone) enhancing sensitivity.

There are several proposals to attempt the first observation of this effect for NR. In~\cite{nakamura21}, a position-sensitive gaseous detector is considered to identify two clusters, one due to the NR and the other produced by a separated de-excitation X-ray. The detector is a quite small (27~l volume) gas TPC with micro-pattern and electroluminescence readouts and filled with Ar (at 1~atm) or Xe (at 8~atm). The neutron beam to produce the recoils has an energy of 565~keV and comes from the $^7$Li(p,n)$^7$Be reaction. According to a Monte Carlo study, in these conditions the Migdal effect signal (at a rate of 10$^2$-10$^3$~c$/$day) would be observable provided backgrounds are controlled. In the proposal by the MIGDAL (Migdal In Galactic Dark mAtter expLoration) experiment~\cite{migdal}, the goal is to measure the ER and NR from a same vertex in a low pressure gaseous detector using high neutron fluxes from D-D and D-T generators (producing neutron energies of 2.45 and 14.1~MeV, respectively). An optical TPC filled with gas CF$_4$ (pure or mixed with Xe or Ar) at 50~Torr and equipped with glass-GEMs is being considered; the charge and light readout with a CMOS camera and a photomultiplier allows 3D reconstruction for few mm long tracks. Simulation and experimental work is underway to quantify signal and background and set-up the experiment at the Rutherford Appleton Laboratory in the UK.

\section{Techniques and results}
\label{sectech}

Many different and complementary techniques are being applied or under consideration in the direct detection of DM based on the measurement of the produced charge, light or heat signals or, in hybrid detectors, of a combination of two of them~\cite{schumann2019,appecdm}. In this section, the main techniques will be briefly presented and the latest results obtained in the context of different experiments for each of them will be discussed.

\subsection{Liquid Ar and Xe detectors}

Noble gases like Xe or Ar scintillate and can be ionized. In liquid state, they can form massive DM targets, which have produced leading results in the high mass range from a few GeV to TeV. A dual-phase TPC can measure both the primary scintillation (S1) and the secondary scintillation from drifted electrons (S2); the ratio of these signals allows to distinguish ER and NR, with an energy threshold at the level of 1~keV$_{ee}$\footnote{Electron-equivalent energy.}. The consideration of drift times allows 3D event reconstruction typically at mm scale.

Three Xe experiments have released over the last years the strongest constraints on WIMP-nucleus interaction above 6~GeV: XENON1T at Laboratori Nazionali del Gran Sasso (LNGS) in Italy~\cite{Aprile:2018dbl}, LUX at the Sanford Underground Research Facility (SURF) in US~\cite{Akerib:2016vxi} and PANDAX-II at Jinping Underground Laboratory in China~\cite{Cui:2017nnn,pandaxiifull}. Considering only the S2 signal, above 0.4~keV$_{ee}$, XENON1T has also derived results for electron scattering~\cite{Aprile:2019xxb}. Searches for light DM, considering the Migdal effect and Bremsstrahlung looking for ER, have been performed by XENON1T~\cite{Aprile:2019jmx} and LUX~\cite{Akerib:2018hck}. Many different DM models have been taken into account in the analyses (inelastic scattering, effective interactions, bosonic Super-WIMPs, light mediators, dark photons $\dots$); axions have been proposed as an explanation of the excess of ER observed by XENON1T~\cite{Aprile:2020tmw}. Extensions of these experiments with several tonnes of active mass have been built for operation in the first half of this decade: LUX-ZEPLIN (LZ) is being commissioned at SURF, XENONnT is taking data at LNGS and PANDAX-4T at Jinping has already released first results~\cite{pandaX4T} with a lowest excluded cross section (90\% C.L.) of 3.8$\times$10$^{-47}$~cm$^2$ at a DM mass of 40~GeV. In a longer term, the DARWIN observatory is expected to start at the end of the decade using 40~tonnes.

In Ar detectors, the different scintillation pulse shape of ER and NR provides a very efficient discrimination method. DEAP-3600 is a single-phase detector operated at SNOLAB in Canada that has shown an outstanding background rejection factor~\cite{Ajaj:2019jk}. DarkSide-50, at LNGS, has used a dual-phase detector filled with underground Ar (UAr) having a reduced (by a factor 1400$\pm$200) $^{39}$Ar content~\cite{Agnes:2018ep}; a low mass DM search, detecting S2 only with a 100~eV$_{ee}$ threshold, produced leading sensitivity at 1.8–3.5~GeV~\cite{Agnes:2018fg} and results on electron scattering~\cite{Agnes:2018ft}. The Global Argon DM Collaboration (GADMC) has been formed to work on the development of SiPMs and on the procurement of ultra-pure UAr extracted from the Urania plant in Colorado (US) and purified at the Aria facility in Sardinia (Italy)~\cite{aria}. The DarkSide-20k experiment at LNGS will be the first step, with a smaller version using a 1~tonne target optimized for low mass DM searches (DarkSide-LowMass). The DArTInArDM detector~\cite{dart} has been designed to verify the low activities of $^{39}$Ar required in the UAr and the Recoil Directionality (ReD) detector is in development too exploring columnar recombination~\cite{red}. For the end of the decade, the ARGO detector with 360~tonnes (fiducial mass) is foreseen at SNOLAB.

\subsection{Bolometers}

In solid-state cryogenic detectors, phonons are measured by the tiny temperature increase induced, which requires operation at temperatures of tens of mK at most; a simultaneous measurement of ionization or scintillation allows discrimination of ER and NR. Very low mass crystals have reached low thresholds even below 100~eV, and therefore this type of detectors are producing leading results in GeV and sub-GeV regions.

EDELWEISS-III operated at the Laboratoire Souterrain de Modane (LSM) in France 24~Ge bolometers (870~g each), presenting very good results at 5-30~GeV and limits also on Axion-Like Particles (ALPs)~\cite{Hehn:2016nll}. For the EDELWEISS-subGeV program, much smaller Ge bolometers (33~g) are being used now, having achieved a 60~eV threshold to explore DM masses down to 45~MeV (considering the Migdal effect) and electron interactions and dark photons~\cite{Armengaud:2019kfj,Arnaud:2020svb}. New searches via Migdal effect obtained with a 200~g cryogenic Ge  detector with NbSi TES sensors in Modane have allowed to go down to masses of 32~MeV~\cite{edelmigdal}.

SuperCDMS operated Ge and Si bolometers of hundreds of grams at the Soudan Underground Laboratory in US. A 70~eV threshold was achieved exploiting the Neganov-Trofimov-Luke (NTL) effect at high bias voltage (HV) to convert charge into heat; results for masses down to 1.5~GeV were presented from different analyses~\cite{Agnese:2017njq,Agnese:2018gze}.  Data have been reanalyzed considering Bremsstrahlung radiation and the Migdal Effect, which allows to explore masses down to 30~MeV~\cite{scdmsmigdal}. Very small Si detectors (0.93 and 10.6~g) have been operated on surface considering nucleon and electron scattering and dark photons~\cite{Amaral:2020ryn,Alkhatib:2020slm}. The operation of SuperCDMS at SNOLAB in Canada with different types of Ge and Si detectors (up to $\sim$30~kg) is expected to start soon.

CRESST-III is operating at LNGS 10~CaWO$_4$ scintillating bolometers (24~g each), after using larger crystals in CRESST-II~\cite{Angloher:2015ewa}. A 30~eV threshold has been achieved, producing the best limits for WIMP-nucleus SI interaction down to 160~MeV and also for SD scattering (neutron-only case) on $^{17}$O~\cite{Abdelhameed:2019hmk}. As it is possible to include other nuclides in the target material, SD interactions (both proton- and neutron-only cases) on $^7$Li~\cite{cresstsd2019} and on $^6$Li ~\cite{cresstsd2022} have been studied above ground too. The use of up to 100 crystals is foreseen, with the goal of lowering the threshold to 10~eV.

\subsection{Ionization Ge and Si detectors}

Purely ionization detectors, like those based on semiconductors, are also producing very interesting results in the search for low mass DM.

Point-Contact Ge detectors can reach sub-keV thresholds thanks to a very small capacitance in combination with a large target mass, as shown by the CoGENT detector at Soudan~\cite{cogent}. Detectors with a mass of $\sim$1~kg each and a threshold of 160~eV$_{ee}$ are being operated at Jinping by CDEX (China Dark matter EXperiment); CDEX-1, with one detector, set constraints on WIMP-nucleon SI and SD couplings~\cite{cdex1} also including the Migdal effect~\cite{Liu:2019kzq}. Limits from an annual modulation analysis have been presented~\cite{cdexmod}. CDEX-10, using a 10~kg detector array immersed in liquid N$_2$ has already released results~\cite{cdex10} and larger set-ups are in preparation for CDEX-100 and CDEX-1T.

In Silicon charge-coupled devices (CCDs), the charge produced in the interaction drifts towards the pixel gates, until readout; they offer 3D position reconstruction and effective particle identification for background rejection although have a long signal collection time. DAMIC (DArk Matter In CCDs), operating 7~CCDs (6~g each) at SNOLAB with a threshold 50~eV$_{ee}$, has presented results on electron scattering and hidden photon DM~\cite{damic1,damic2} and on nucleon scattering~\cite{damic3}. DAMIC-M will be operated at LSM using 50 larger CCDs (13.5~g each) with Skipper readout, where the multiple measurement of the pixel charge allows to reduce noise and achieve single electron counting with high resolution. This technology is already being used in SENSEI (Sub-Electron-Noise Skipper CCD Experimental Instrument); small prototypes (0.0947 and 2~g) have been operated at Fermilab, setting leading constraints on DM-electron scattering and hidden-sector candidates~\cite{sensei,sensei2}. SENSEI plans to install a 100-g detector with 48~CCDs at SNOLAB. The project for the end of the decade is OSCURA, using 10~kg of target.

\subsection{Gaseous charge detectors}
Gaseous detectors based on pure ionization are also in development focused on low mass DM. A Spherical Proportional Counter (SPC) is able to achieve very low threshold thanks to a very low capacitance for a large volume~\cite{giomataris}, being the anode a small ball at the center of the sphere. The SEDINE detector, consisting of a copper sphere, 60~cm in diameter, filled with Ne-CH$_{4}$ at 3.1~bar (280~g active mass) was operated at LSM. Exclusion results for WIMPs were derived from a 42~day run at 50~eV$_{ee}$ acquisition threshold~\cite{newsg1}, as well as limits on the axion-photon coupling for solar Kaluza-Klein axions~\cite{newsgaxions}. NEWS-G (New Experiments With Spheres-Gas) has built a new, larger copper sphere, 140~cm in diameter. After some tests, commissioning data with CH$_4$ (at 135~mbar) were taken at LSM and now the sphere is at SNOLab for operation. The single electron response (gain, drift or diffusion times) has been characterized with a laser system~\cite{newsg2} and efforts have been made to mitigate the effect of $^{210}$Pb in copper~\cite{Balogh}. The first measurement of the quenching factor for Ne has been achieved deploying a SPC in a neutron beam at the Triangle Universities Nuclear Laboratory (TUNL, Duke University)~\cite{qne}. The development of spheres made of electroformed Cu (ECUME, DARKSPHERE) is underway in collaboration with Pacific Northwest National Laboratory (PNNL).

Other gaseous detectors are being considered, like a high pressure TPC equipped with Micromegas (MicroMesh Gas Structures) readouts, offering low intrinsic radioactivity, background discrimination based on the event topology and the possibility of scaling-up. TREX-DM (Tpcs for Rare Event eXperiments-Dark Matter) is based on a TPC holding a pressurized gas at 10~bar inside a copper vessel, equipped with the largest Microbulk Micromegas readouts (surface of 25$\times$25~cm$^2$) ever built~\cite{trexdmiguaz,trexdmigor,trexdmbkg}. It operates at the Canfranc Underground Laboratory (LSC) in Spain, being presently at the commissioning phase.
The target is flexible and the available volume (24~l) corresponds to $\sim$0.3~kg of Argon and $\sim$0.16~kg of Neon at 10~b. Runs with mixtures of both Argon and Neon with isobutane (atmospheric Ar+1\%iC$_{4}$H$_{10}$ and Ne+2\%iC$_{4}$H$_{10}$) up to 8~bar have been made. The prospects for the energy threshold are from 0.4~keV$_{ee}$ down to 0.1~keV$_{ee}$ and the background model developed points to levels at 1 to 10~keV$^{-1}$ kg$^{-1}$ day$^{-1}$ in the region of interest~\cite{trexdmbkg}.

\subsection{Bubble chambers}

Bubble chambers use target liquids kept in metastable superheated state so that sufficiently dense energy depositions start the formation of bubbles, which are read by cameras. Although with this technique no direct measurement of the recoil energy is obtained, they are almost immune to ER background sources and, as most of the considered targets contain $^{19}$F, they offer the highest sensitivity to SD proton couplings. 

Merging PICASSO and COUPP collaborations, PICO has operated a series of bubble chambers at SNOLAB. In PICO-60, with 52~kg of C$_{3}$F$_{8}$ a 2.45~keV threshold for nuclear recoils was achieved and the best SD WIMP-proton limit from direct detection has been derived~\cite{pico}; results on photon-mediated DM-nucleus interactions have been recently derived too \cite{pico2}. Some changes in the design, like the buffer-free concept~\cite{bressler}, have been implemented in PICO-40L, already starting the data taking. PICO-500 is a fully funded tonne-scale chamber now in design phase.

\subsection{Scintillating crystals}

NaI(Tl) crystals read by photomultipliers (PMTs) are being used in projects focused on the detection of the DM annual modulation, since, being quite cheap and very robust detectors, it is possible to accumulate a large target mass and to run for long times in very stable conditions. New developments have been necessary to get an ultra-low background (reducing for instance $^{40}$K and $^{210}$Pb activity in the crystals) and a low energy threshold. 

The DAMA/LIBRA experiment, using $\sim$250~kg of NaI(Tl) scintillators at LNGS, is collecting data for more than twenty years with modules (crystal mass of 9.7~kg each) produced by Saint Gobain company. The software energy threshold was reduced from 2 to 1~keV$_{ee}$ in the second phase of the experiment using new PMTs since 2011. The results of the two phases~\cite{damaphase1,damaphase2} favour the presence of a modulation with all the proper features for DM; from the whole analyzed exposure of 2.86~t$\cdot$y, the C.L. of this signal is at 13.7$\sigma$ and the deduced modulation amplitude for the 2-6~keV$_{ee}$ region is $S_{m}=(0.01014\pm0.00074)$~cpd/kg/keV \cite{dama2021}; compatible values have been found for different fitting procedures, periods of time, energy regions from 1 to 6~keV$_{ee}$ and detector units. Improved model-dependent corollary analyses after DAMA/LIBRA phase 2 have been presented~\cite{damacorollary}, applying a maximum likelihood procedure to derive allowed regions in the parameters' space of many different considered scenarios by comparing the measured annual modulation amplitude with the theoretical expectation.
 
\begin{figure}[]
\includegraphics[width=22pc]{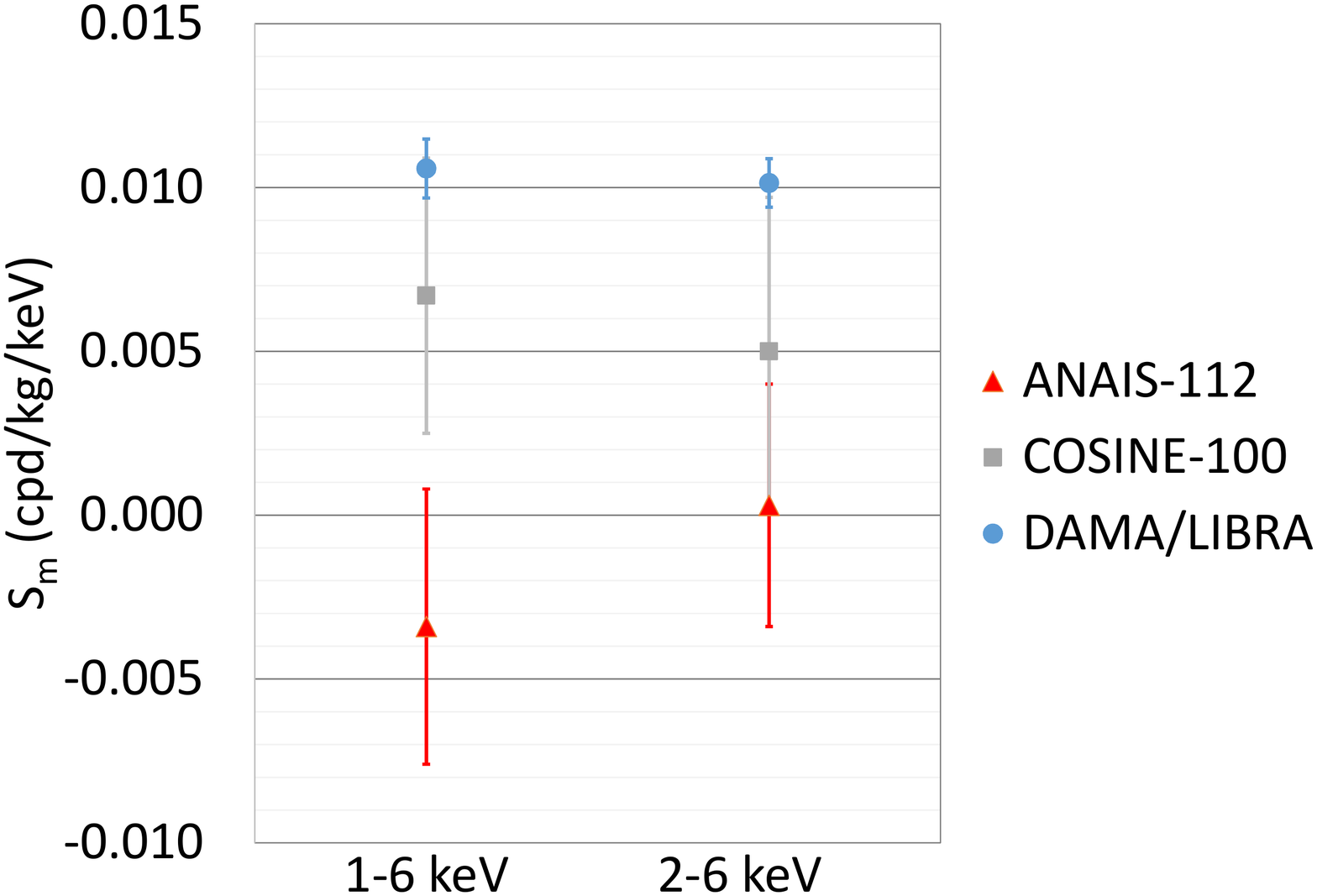}\hspace{2pc}%
\begin{minipage}[b]{14pc}\caption{\label{plotam} Modulation amplitudes obtained by DAMA/LIBRA~\cite{dama2021}, ANAIS-112~\cite{anais2021} and COSINE-100~\cite{cosine2021} experiments in the energy regions of 1-6~keV$_{ee}$ and 2-6~keV$_{ee}$.}
\end{minipage}
\end{figure}

However, following the exclusion results presented by many experiments~\cite{appecdm}, there is strong tension when interpreting DAMA/LIBRA results as due to DM, not only in the standard paradigm but even assuming more general halo and interaction models. In addition, no annual modulation signal has been found with different targets like Xe~\cite{xenon100,xmass,lux} or Ge~\cite{cdexmod}. In this context, a model-independent proof or disproof with the same NaI target was considered mandatory, and this is the goal of several projects all over the world, being ANAIS-112 and COSINE-100 now in data-taking phase. Figure~\ref{plotam} compares the modulation amplitudes derived by the three experiments.

ANAIS (Annual modulation with NAI Scintillators) is operating at LSC 9 NaI(Tl) modules (112.5~kg in total) built by Alpha Spectra company~\cite{anaisperformance,anaisbkg,anaissensitivity}. Data taking started in 2017, with an outstanding light collection of $\sim$15 phe/keV for all modules allowing an energy threshold at 1~keV$_{ee}$. After the first results released on model-independent annual modulation for 1.5~y and exposure of 157.55~kg$\cdot$y ~\cite{anaismod} supporting the null hypothesis, the analysis of 3~y for 313.95~kg$\cdot$y has been published~\cite{anais2021}, finding for the 2-6~keV$_{ee}$ region $S_{m}=(0.0003\pm0.0037)$~cpd/kg/keV and for 1-6~keV$_{ee}$ $S_{m}=(-0.0034\pm0.0042)$~cpd/kg/keV, incompatible at 2.7$\sigma$ and 3.3$\sigma$ respectively with DAMA/LIBRA results and confirming the possibility to explore the 3$\sigma$ DAMA/LIBRA region in 5~y.

In COSINE-100, 8 NaI(Tl) detectors also from Alpha Spectra (106~kg in total) are operated immersed in liquid scintillator at the Yangyang underground Laboratory in South Korea~\cite{cosineperformance,cosinebkg} since 2016. The first annual modulation analysis corresponding to 1.7~y with $\sim$61~kg~\cite{cosinemod} have been updated for 2.8~y, with the best fit modulation amplitudes $S_{m}=(0.0050\pm0.0047)$~cpd/kg/keV for 2-6~keV$_{ee}$ and $S_{m}=(0.0067\pm0.0042)$~cpd/kg/keV for 1-6~keV$_{ee}$~\cite{cosine2021}, compatible with both null hypothesis and the DAMA/LIBRA signal. Additionally, strong constraints on the DM interpretation of DAMA/LIBRA signal in the context of different models have been derived~\cite{cosineexc}.

Other NaI projects are in preparation. SABRE (Sodium-iodide with Active Background REjection) plans to operate $\sim$50~kg and set identical detectors in northern and southern hemispheres (at LNGS and Stawell Laboratory in Australia)~\cite{sabre} to investigate any seasonal effect due to backgrounds, which should show opposite phase. They are focused on the development of ultra-high purity NaI(Tl) crystals and tests with one 3.5~kg detector (SABRE Proof of Principle, PoP) are underway at LNGS~\cite{sabrecrys}. The PICOLON project (Pure Inorganic Crystal Observatory for LOw-energy Neutr(al)ino) is working in Kamioka in Japan to produce also radiopure NaI(Tl) scintillators~\cite{picolon} while COSINUS (Cryogenic Observatory for SIgnatures seen in Next-generation Underground Searches) is developing at LNGS NaI scintillating bolometers with capability to discriminate ER and NR~\cite{cosinus}.

\subsection{Directional detectors}

There are two different approaches for directional detectors: the use of nuclear emulsions and the operation of low pressure ($\sim$0.1~atm) gaseous targets in TPCs with different electron amplification devices and track readouts, like Multi-Wire Proportional Chambers (MWPC), Micro Pattern Gaseous Detectors (MPGDs) and optical readouts~\cite{battat,recimaging}. Gas detectors are mostly based on mixtures with $^{19}$F. Table~\ref{summarytable} summarizes the main properties of some of the projects in the field. 

\begin{center}
\begin{table}[h]
\centering
\caption{\label{summarytable} Main properties of some of the projects developing directional detectors.}
\begin{tabular}{lcccc}
\br
Experiment & Technique & Laboratory & Size & Reference \\
\mr
DRIFT & TPC+MWPC & Boulky (UK) &  1 m$^3$ & \cite{battat2,battat3}\\ 
MIMAC & TPC+Micromegas & LSM (France) & 1 m$^3$ (in prep) & \cite{tao,tao2}\\
NEWAGE & TPC+$\mu$PIC & Kamioka (Japan) & 31$\times$31 cm$^2$, 41 cm drift & \cite{yakabe,ikeda}\\ 
DMTPC & TPC+optical readout & WIPP (US) & 1 m$^3$ (in prep) & \cite{deaconu}\\
CYGNO & TPC+optical readout & LNGS (Italy) & 1 m$^3$ (in prep) & \cite{baracchini,domingues}\\
\mr
NEWS-dm & Emulsion+optical readout & LNGS (Italy) & 10 g (prototype) & \cite{emulsion,newsdm}\\
\br
\end{tabular}
\end{table}
\end{center}

An emulsion film made of silver halide crystals dispersed in a polymer can act as target and tracking detector; nuclear recoils produce nm-sized silver clusters and 3D tracks are reconstructed with an optical microscope~\cite{emulsion}. This is the approach followed by NEWSdm (Nuclear Emulsions for WIMP Search with directional measurement)~\cite{newsdm}. New generation nuclear emulsions with nanometric grains (NIT (Nano Imaging Tracker) emulsions) together with fully automated scanning systems overcoming diffraction limits have been developed; a spatial resolution of 10~nm has been achieved. Test at LNGS with 10~g are underway to assess backgrounds. They propose for a Physics run with 10~kg-year a detector placed on an equatorial telescope (to absorb Earth rotation) to keep orientation towards the Cygnus constellation.

DRIFT (Directional Recoil Identification From Tracks) was the pioneer of directional detectors, using MWPCs attached to a TPC with a large conversion volume (1~m$^{3}$) filled with electronegative gas; in this way, the formed ions (not electrons) are drifted to the readout, to reduce diffusion and optimize track resolution. The approach is scalable to large volumes although with limited spatial granularity ($>$1~mm). DRIFT operated at the Boulby Underground Laboratory in UK, over more than a decade, using a CS$_{2}$+CF$_{4}$+O$_{2}$ mixture (0.140~kg at 55~mbar). They measured directional nuclear recoils (from $^{252}$Cf neutrons) quantifying the head-tail asymmetry parameter~\cite{battat2} and derived limits for SD WIMP-proton interaction from 54.7~live-days~\cite{battat3}.

MIMAC (MIcro-tpc MAtrix of Chambers) also operates a dual TPC with a common cathode, but equipped with pixelized bulk Micromegas, at the Modane Underground Laboratory since 2012. They worked with CHF$_{3}$(28\%)+CF$_{4}$(70\%)+C$_{4}$H$_{10}$(2\%) at 50~mbar inside a 25~cm long TPC with 10$\times$10~cm$^2$ readout and a stainless steel vessel. 3D tracks of radon progeny nuclear recoils have been registered \cite{riffard}. First observation of $^{19}$F ion tracks at ion beam facilities with angular resolution better than 15$^{o}$ has been reported~\cite{tao,tao2} and quenching factors of He and F with an ion source in Grenoble have been measured. A detector at the scale of 1~m$^{3}$ is in preparation.

NEWAGE (NEw generation WIMP search with an Advanced Gaseous tracker Experiment) uses a system with amplification structure and readout in a monolithic detector with a TPC and a micro-pixel chamber ($\mu$-PIC). After first operation in surface, long runs at Kamioka have been made using CF$_{4}$ at 100~mb. They have also confirmed the head/tail effect above 100~keV and presented results for SD interaction ~\cite{nakamura15,yakabe}. A low background detector with polyimide is running since 2018 and new limits have been presented \cite{hashimoto,ikeda}. Operation with a different gas, SF$_{6}$, is being tested~\cite{ikeda2}.

DMTPC (Dark Matter Time-Projection Chamber) is based on a TPC equipped with external optical and charge readouts. CCDs offer high granularity, low cost and easy acquisition. Several prototypes (with 10 and 20~l) have been developed since 2007, operated first at MIT and then underground at the WIPP facility in US, using CF$_{4}$ gas at low pressure (30-100~Torr). The measurement of the direction of recoils has been reported too~\cite{deaconu}. A detector at the scale of 1~m$^{3}$ is in preparation using four TPCs (1.2~m in diameter and 27~cm in length).

Gathering most of the groups working on directional dark matter detection, the CYGNUS collaboration was formed to analyze different gas mixtures (with low and high density, considering CF$_{4}$ and SF$_{6}$), to reduce the energy threshold even below 1~keV$_{ee}$ in these detectors and enlarge the volumes (from 10 up to 1000~m$^3$)~\cite{cygnus}. The goal is to develop a multi-site, multi-target observatory at the ton scale for DM and solar neutrino with directionality. There are proposals for CYGNUS detectors in different labs in Australia, Italy, Japan, UK and US. The CYGNO/INITIUM apparatus, at LNGS, will be a 1~m$^{3}$ symmetric detector filled with a mixture of He and CF$_{4}$ (60/40, $\sim$1 kg); each volume will be equipped with triple-GEM structure and read by a sCMOS sensor and a fast light detector (PMT or SiPM)~\cite{costa,baracchini,domingues}. Various prototypes have been built and operated. LEMOn used He/CF$_{4}$ 1~atm (7~l, with 20~cm drift), achieving an energy threshold at 1~keV$_{ee}$ and showing stability over one month \cite{lemon}; nuclear recoils from a neutron gun with direction and sense visible have been registered too. The LIME detector (55~l, with 50~cm drift) has been already tested above ground with a 0.5~keV$_{ee}$ threshold and will operate at LNGS \cite{lime}.

\subsection{New technologies}
Many new ideas to directly detect DM particles have been proposed and some of them are at R\&D phase aimed at measuring smaller and smaller energy depositions and improving the background suppression. Some of them are briefly described in this section. 

\begin{itemize}
\item Scintillating bubble chambers combine the advantages of a bubble chamber (extreme electron rejection and simple instrumentation) with the event-by-event energy resolution of a liquid scintillator. This technique has been established for a 30~g xenon bubble chamber~\cite{baxter} and is being developed by the SBC Collaboration, preparing 10~l LAr chambers at Fermi National Laboratory to operate at SNOLAB in coming years; the goal is to have 100~eV threshold and 10~kg$\cdot$yr exposure.
\item In the SnowBall proposal, supercooled water is used so that an incoming particle triggers crystallization of purified water using a camera for image acquisition; the technique, insensitive to ER, has been tested with neutrons at $-$20$^{o}$~\cite{snowball}.
\item The HeRALD project (Helium Roton Apparatus for Light Dark Matter) proposes the use of superfluid $^{4}$He as target and low temperature calorimeters (TES) as sensors to measure photons and quasiparticles by quantum evaporation~\cite{herald}. Single phonon detection for DM via quantum evaporation and sensing of $^{3}$He has been proposed too~\cite{evaphe3}.
\item High purity lab-grown diamond crystals as target outfitted with cryogenic phonon and charge readout could be sensitive to DM scattering of very low mass candidates~\cite{kurinsky}. Carbon is lighter than other semiconductor materials used and low noise allowing a sub-eV threshold is expected. Results from the first cryogenic detector with diamond as absorber have been recently presented, reaching an energy threshold as low as 16.8~eV~\cite{diamondcresst}.
\item In the so-called paleo-detectors, persistent traces left by DM interaction in ancient minerals could be searched for profiting from a large integration time~\cite{paleo,paleo2,paleo3,paleo4,paleo5}. Different readout scenarios are considered to achieve nm resolution.
\end{itemize}

New ideas have arisen specifically to obtain the recoil direction:
\begin{itemize}
\item Crystal defect spectroscopy could be used as WIMP-induced NR in a diamond target would create
an observable damage trail that alters the strain pattern. The large target mass of a solid-state detector would be combined with an ultra-fine spatial resolution (at nm-scale)~\cite{rajendran,marshall,ebadi}.
\item In a DNA strand detector, DNA strands mounted onto a nm-thick gold foil could be
severed by a recoiling gold atom kicked out by a WIMP; biological techniques could allow to identify position of each severing event~\cite{dna,ohare}.
\item Planar targets, like graphene, make easier the determination of recoil direction by avoiding multiple interaction in bulk targets~\cite{wang}. In PTOLEMY-G graphene FETs with tunable meV band gaps are proposed~\cite{graphene}.
\item Anisotropic scintillators made of ZnWO$_4$ or organic compounds are being studied too~\cite{pedersen,blanco}. In~\cite{directani}, the directional detectability of DM with single phonon excitations has been compared for 26 anisotropic targets by analyzing the expected daily modulation.
\end{itemize}

\section{Summary and Outlook} \label{secsum}

The direct detection of DM particles is really challenging due to the small and rare signal expected. Complementary experiments, based on different detection technologies and targets, are ongoing exploring different interactions, mass ranges of candidates and possible signatures.

For SI WIMP-nucleon interactions, above 10~GeV the best limits on cross section come presently from xenon experiments. At lower masses and for SD interactions, results from different types of detectors give the stringent constraints: cryogenic detectors, liquid noble detectors operated for charge collection only, purely ionization detectors and bubble chambers; the use of light targets achieving extremely low energy thresholds and/or searching for different interaction channels are the strategies followed and new detection technologies are in development. The proposed Migdal effect helps to enhance sensitivity particularly for low mass DM. In addition, important results (even if still with low significance) from NaI(Tl) experiments have been presented to solve the long standing conundrum of the DAMA/LIBRA annual modulation result. Concerning WIMP-electron scattering, SENSEI and XENON have presented the lowest limits for the interaction cross section below and above $\sim$10~MeV, respectively. 

TPC-related experiments and projects are giving very important contributions to the direct detection of DM. Liquid noble detectors with Xe and Ar operated as dual-phase TPCs have provided leading constraints for SI WIMP-nucleus interaction, down to 1.8~GeV when operated in S2-only mode, and the huge projects like DARWIN and ARGO are expected to explore cross sections down to the irreducible neutrino background. Gaseous detectors with light targets like Ne and sub-keV threshold have very good prospects for low mass DM. Low-pressure TPCs with different readouts (MWPCs, Micromegas, $\mu$PICs, optical CCDs, \dots) are ideally suited for directional DM detectors, essential to prove the galactic origin of a possible signal; the already built detectors (with volumes from 0.1 to 1~m$^{3}$) have measured directional NR, confirmed the head-tail effect and already set limits for SD WIMP-proton interaction. Although imaging short tracks with sufficient resolution and sampling large volumes is hard, significant progress on basic requirements (like radiopurity, homogeneity, stability and scalability) is being made. Finally, it is worth mentioning that low-pressure optical TPCs or position-sensitive gaseous detectors are being considered to attempt the first observation of the Midgal effect of NR.


\end{document}